\documentclass[reqno,11pt]{amsart}
\usepackage{multind}
\usepackage{hyperref}
\usepackage{graphicx}
\usepackage{amscd}
\usepackage[mathscr]{eucal}
\numberwithin{equation}{section}
\allowdisplaybreaks[1]

\newcommand{\SetFigFont}[3]{}

\title[Error Estimates for the Riccati Equation]{Error Estimates for Approximate Solutions
of the Riccati Equation with Real or Complex Potentials}

\author[F.\ Finster]{Felix Finster}
\thanks{F.F.\ is supported in part by the Deutsche Forschungsgemeinschaft.}
\address{NWF I - Mathematik \\ Universit\"at Regensburg \\ D-93040 Regensburg \\ Germany}
\email{Felix.Finster@mathematik.uni-regensburg.de}

\author[J.\ Smoller]{Joel Smoller \\ \\ July 2008}
\thanks{J.S.\ is supported in part by the Humboldt Foundation and the National Science Foundation,
Grant No.~DMS-0603754.}
\address{Mathematics Department \\ The University of Michigan \\ Ann Arbor, MI 48109, USA}
\email{smoller@umich.edu}

\newtheorem{Def}{Def.}[section]
\newtheorem{Thm}[Def]{Theorem}

\newtheorem{Lemma}[Def]{Lemma}

\newtheorem{Example}[Def]{Example}
\newcommand{\Proof}{\begin{proof}}
\newcommand{\QED}{\end{proof} \noindent}
\newcommand{\QEDrem}{\ \hfill $\Diamond$}
\newcommand{\spc}{\;\;\;\;\;\;\;\;\;\;}

\newcommand{\C}{\mathbb{C}}
\newcommand{\R}{\mathbb{R}}

\newcommand{\beq}{\begin{equation}}
\newcommand{\eeq}{\end{equation}}

\begin{document}
\maketitle

\begin{abstract}
A method is presented for obtaining rigorous error estimates for approximate solutions of
the Riccati equation, with real or complex
potentials. Our main tool is to derive invariant region estimates for complex solutions
of the Riccati equation. We explain the general strategy for applying these estimates
and illustrate the method in typical examples, where the approximate solutions are
obtained by glueing together WKB and Airy solutions of corresponding
one-dimensional Schr\"odinger equations.
Our method is motivated by and has applications to the analysis of linear wave equations
in the geometry of a rotating black hole.
\end{abstract}

\section{Introduction}
\setcounter{equation}{0}
The time-independent one-dimensional Schr\"odinger equation plays an important role in both mathematics
and physics. Except in special situations, it is impossible to find exact
solutions, and thus one resorts to approximation techniques.
The most common method is to take the WKB wave functions away from the zeros of the potential,
and to glue them together with Airy functions, which approximate the solutions near the
zeros (see for example the textbook~\cite[Section~2.4]{Sakurai} or the book~\cite{FF}).
Usually, the potential in the Schr\"odinger equation is real-valued. However, there
are situations (like systems with dissipation) where
the Schr\"odinger potential is complex. 

The shortcoming of most approximation procedures is that they do not give error estimates.
To quote Fr\"oman and Fr\"oman \cite[page 3]{FF}~(1965),
\begin{quote}
``In spite of the abundant literature there is still lacking a convenient method for obtaining definite
limits of error\ldots The problem of obtaining such limits of error is not only of academic or
mathematical interest but is especially important in some physical applications\ldots''
\end{quote}
Unfortunately, such a convenient method for obtaining definite limits of error is still lacking today.
Indeed, to our knowledge, despite the vast literature on WKB methods (see~\cite{FGF} and the references therein),
only a few papers are concerned with rigorous error estimates.
As typical examples see~\cite{Maslov, Molzahn},
where $L^2$-estimates are derived in the time-dependent setting for short times, as well as~\cite{JMS},
where estimates are obtained in the setting of multiple tunneling.
Here we shall develop a general method for obtaining {\em{rigorous}}
pointwise error estimates for any given approximate solution in the time-independent setting,
even in the case when the Schr\"odinger potential is complex.

We were led to the study of error estimates for solutions of one-dimensional Schr\"o\-din\-ger equations
from problems arising in general relativity. In particular, when analyzing the scalar wave equation in
the Kerr geometry, Carter's remarkable separation of variables~\cite{C00} gives rise to a
radial and angular ODE. These equations can both be transformed into the Schr\"odinger form,
which are coupled through two parameters~$\omega$ and~$\lambda$.
For estimating the global behavior of the solutions uniformly in~$\omega$ and~$\lambda$,
we were led to the idea of using invariant disk estimates~\cite{FS, FKSY2}.
When attempting to extend our analysis to gravitational waves in the framework of the Teukolsky
equation~\cite{Teukolsky}, one of the difficulties that arises is that the potentials in the
separated equations turn out to be complex. In the present paper, we obtain invariant
disk estimates in a general setting, which are applicable to one-dimensional Schr\"odinger
equations with a real or complex potential. The previous estimates in~\cite{FS, FKSY2} will arise as
special cases. In particular, our method gives rigorous pointwise error estimates for the standard
approximations by WKB and Airy functions. But the method works just as well for any other
approximate solutions. The aim of the present paper is to introduce the method from a general
point of view, and to illustrate it in typical examples.
For an application to eigenvalue and gap estimates as well as to problems involving parameters
we refer to~\cite{FS} and~\cite{FKSY2}.

We now set up the framework and fix our notation. The one-dimensional
Schr\"odinger equation reads
\[ \left[ -\frac{d^2}{dx^2} + {\mathcal{V}}(x) \right] \phi(x) \;=\; \lambda\, \phi(x)\:. \]
Absorbing the parameter~$\lambda$ into the potential, we rewrite the Schr\"odinger equation as
\beq \label{schrodinger}
\frac{d^2}{dx^2} \phi \;=\; V(x)\, \phi(x)\:,
\eeq
where the new potential~$V={\mathcal{V}}-\lambda$ may depend on additional parameters.
This is a linear second order ODE. Thus the general solution~$\phi$ can be
written as a linear combination of two linearly independent
fundamental solutions~$\phi_1$ and~$\phi_2$,
\beq \label{gensol}
\phi \;=\; a\: \phi_1 + b\: \phi_2 \:.
\eeq
By choosing the coefficients~$a$ and~$b$ appropriately, one can then satisfy suitable
boundary conditions, if so desired. Setting
\beq \label{ydef}
y \;=\; \frac{\phi'}{\phi}\:,
\eeq
the function~$y$ satisfies the Riccati equation
\begin{equation} \label{riccati}
y'\;=\; V-y^2\:.
\end{equation}
Our strategy is to estimate~$y$ in the {\em{complex plane}}. This requires a brief
explanation, because in the case when~$V$ is real it is natural to
also choose~$\phi$, and thus~$y$, to be real.
This real function~$\phi$ will in general have zeros, and~$y$ will be singular
at each of these zeros. These divergences make it difficult to estimate~$y$.
We avoid this problem by considering a linear combination~$\phi = \phi_1 + i \phi_2$ with
real fundamental solutions~$\phi_1$ and~$\phi_2$. Then, since the zeros of the fundamental
solutions never coincide, the function~$\phi$ never vanishes, and thus the
corresponding function~$y$ is globally well-defined.

In the more general case when the potential~$V$ is complex, the situation is more difficult because
complex solutions of the Riccati equation~\eqref{riccati} may ``blow up in finite time''.
Nevertheless, we can hope to avoid singularities of the Riccati equation by choosing the
initial condition~$y(0)$ suitably.

In this paper, we focus on the analysis of solutions of the Riccati equation
for specific initial conditions~$y(0)$. This is justified because our estimates
also give rise to corresponding estimates for general approximate Schr\"odinger
wave functions, using the following procedure.
Suppose that a specific~$y$ is known, up to a small error term. Then integrating~\eqref{ydef}, one obtains
one fundamental solution~$\phi_1(x) = \exp(\int^x y)$ of the Schr\"odinger equation. The
other fundamental solution can readily be obtained by integrating the Wronskian
equation~$\phi_1' \phi_2 - \phi_1 \phi_2' = {\mbox{const}}$. We thus obtain the general
solution~\eqref{gensol}. Keeping track of the error in this procedure, one can derive error estimates for
any Schr\"odinger wave function.

\section{The Flow of Circles for a Constant Potential} \label{sec2}
\begin{figure}%
\begin{center}%
 \includegraphics[width=7.5cm]{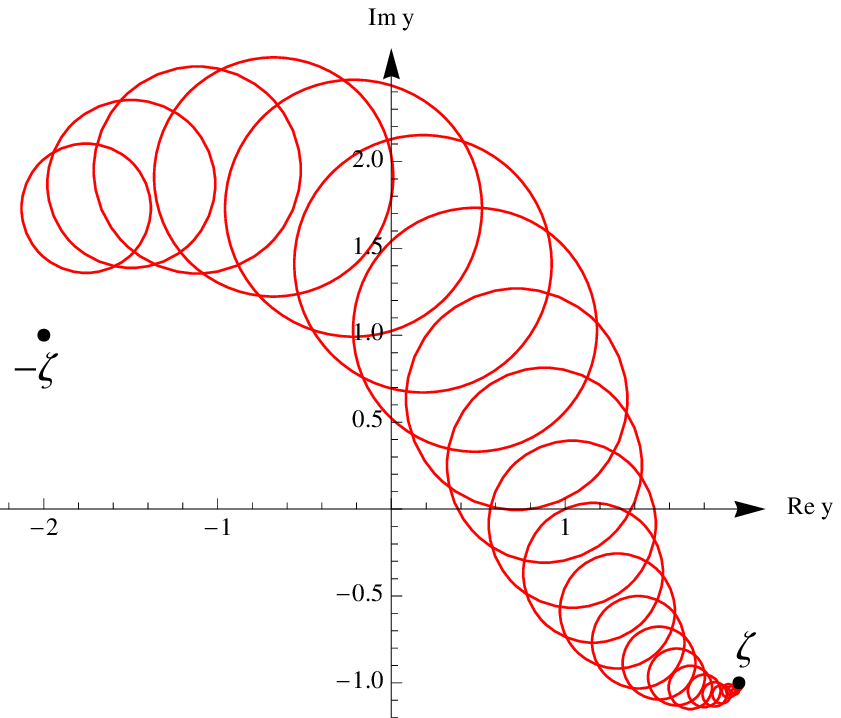}%
  \includegraphics[width=3.7cm]{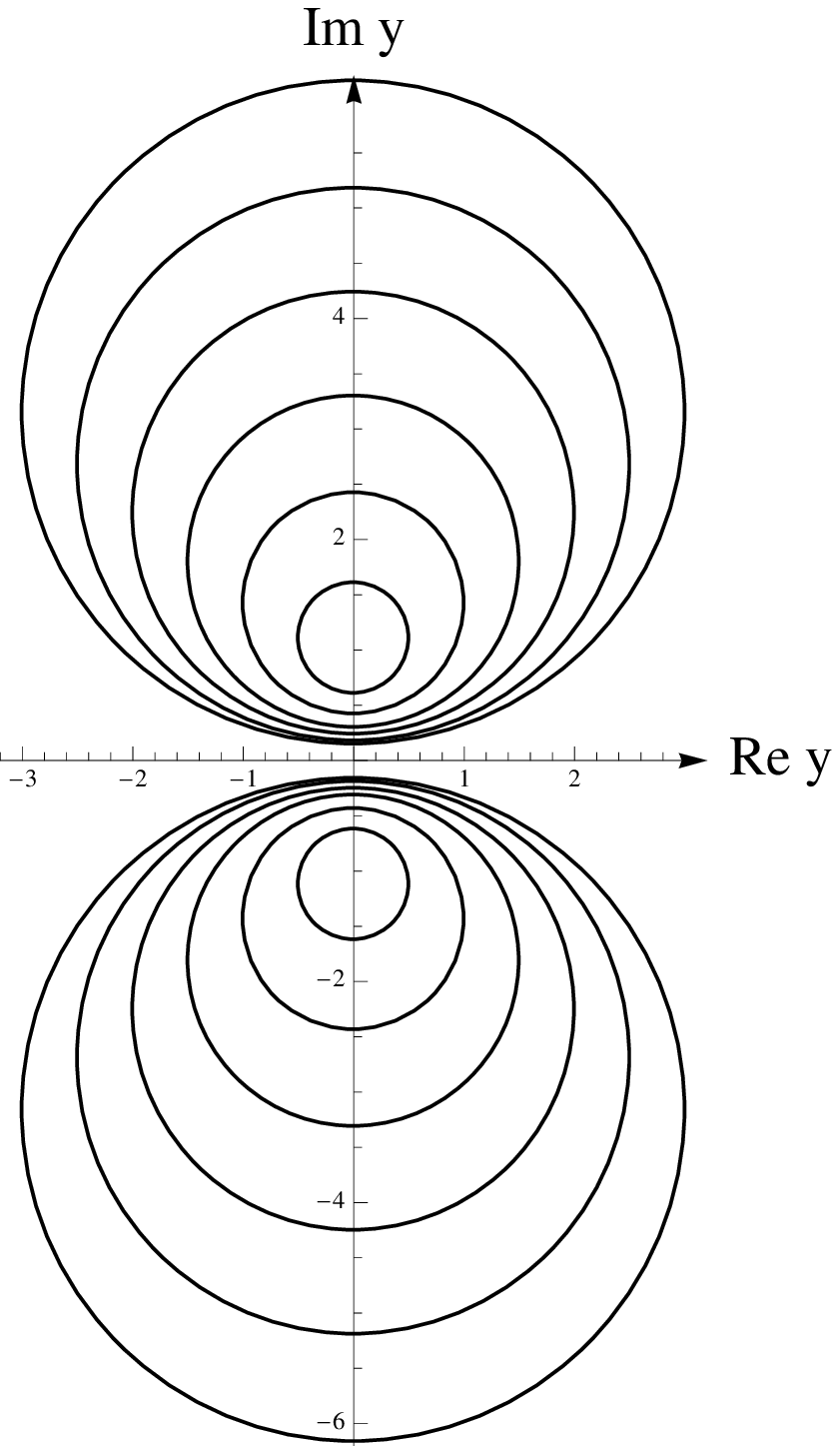}%
\caption{Example for flow of circles with~$\zeta = 2-i$ (left)
and stationary circles in the case~$\zeta = i$ (right).}%
\label{fig0}%
\end{center}%
\end{figure}

\setcounter{equation}{0}
In this section we consider the special case of a constant potential. This case can easily be
analyzed in closed form, and the resulting flow in the complex plane will be helpful
for the geometric understanding of the invariant disk estimates. Thus assume that $V$ is a
non-zero complex constant. We set $\zeta = \sqrt{V}$. Then the general solution of the
Schr\"odinger equation~\eqref{schrodinger} is
\[ \phi(x) \;=\; a e^{\zeta x} + b e^{-\zeta x}\:, \quad x \in \R \]
with $a,b \in \C$. The corresponding solution of the Riccati equation is
\begin{eqnarray*}
y(x) &=& \zeta\: \frac{a e^{\zeta x} - b e^{-\zeta x}}
{a e^{\zeta x} + b e^{-\zeta x}} \;=\; 
\zeta\: \frac{(a-b) \cosh(\zeta x) + (a+b) \sinh(\zeta x)}
{(a+b) \cosh(\zeta x) + (a-b) \sinh(\zeta x)} \\[.2em]
&=& \zeta\: \frac{(a-b) + \tau(x) \,(a+b)}{(a+b)+ \tau(x) \,(a-b)} \:,
\end{eqnarray*}
where $\tau(x):=\tanh(\zeta x)$.
Setting~$y_0 = y(0)$,
\[ y_0 \;=\; \zeta\: \frac{a-b}{a+b} \:, \]
we can express $y$ in terms of $y_0$,
\[ y(x) \;=\; \zeta\:\frac{y_0 + \tau(x) \,\zeta}{\tau(x) \,y_0+\zeta}\:. \]
Note that the mapping~$y_0 \rightarrow y$ is a fractional linear transformation. It has the
fixed points~$y = \pm \zeta$ corresponding to the two solutions~$\phi=e^{\pm \zeta x}$,
respectively. Furthermore, we see that circles
are mapped onto circles or straight lines.
Thus by varying~$x$, we obtain a flow of circles in the complex plane.
To analyze this flow, let us assume that~$y_0$ lies on a circle of radius~$R_0$ around~$m_0$, i.e.
\[ y_0 \;=\; m_0 + z \qquad {\mbox{with}} \qquad z \;=\; R_0\, e^{i \varphi}\:. \]
Then for any fixed~$x$,
\[ y \;=\; \zeta\:\frac{z + m_0 + \tau \zeta}{\tau z + \tau m_0 + \zeta}\:. \]
Thus for any~$m \in \C$, we can write
\begin{eqnarray*}
y-m &=& \frac{(\zeta-\tau m) R_0 e^{i \varphi} + \zeta m_0 + \tau \zeta^2- \tau m m_0 - m \zeta}
{\tau  R_0 e^{i \varphi} + \tau m_0 + \zeta} 
\;=\; \frac{c}{e^{i \varphi}} \left( \frac{e^{i \varphi} + \gamma}{e^{-i \varphi} + \delta} \right) ,
\end{eqnarray*}
where the introduced quantities
\[ c \;=\; \frac{(\zeta-\tau m) R_0}{\tau m_0 + \zeta} \:,\quad
\gamma \;=\; \frac{\zeta m_0 + \tau \zeta^2- \tau m m_0 - m \zeta}{(\zeta-\tau m) R_0}
\:,\quad \delta \;=\; \frac{\tau R_0}{\tau m_0 + \zeta} \]
are independent of~$\varphi$. For~$m$ to be the center of the new circle, the
resulting expression for $|y-m|$ must be independent of~$\varphi$. This gives the
condition~$\gamma=\overline{\delta}$. We thus conclude that
\[ |y-m| \;=\; R \:, \]
where the center and radius are given by
\[ m \;=\; \zeta\: \frac{(m_0 + \tau \zeta)\overline{(\tau m_0 + \zeta)} - R_0^2 \,\overline{\tau}}
{|\tau m_0+ \zeta|^2-R_0^2 |\tau|^2} \:,\qquad
R \;=\; \frac{R_0 \:|\tau^2-1|\: |\zeta|^2}{|\tau m_0 + \zeta|^2-R_0^2 \,|\tau|^2}\:. \]
The denominator can have zeros, in which case the circle degenerates to a straight line.
Whenever the denominator has no zeros, we can take the limit~$x \rightarrow \infty$
to find that in the case~${\mbox{Re}}\, \zeta >0$,
the circle asymptotically approaches the fixed point~$y=\zeta$, unless if we start
at~$m_0=-\zeta$ with~$R_0=0$. Thus we can say that~$y=\zeta$ is the stable
and~$y=-\zeta$ the unstable fixed point. Conversely,
in the case~${\mbox{Re}}\, \zeta <0$, the fixed point~$y=\zeta$ is unstable, whereas~$y=-\zeta$
is stable. The flow of circles is illustrated in Figure~\ref{fig0} (left).

In the remaining case~${\mbox{Re}}\, \zeta = 0$, we clarify the situation by computing
the stationary circles. In this case, $\zeta$ and~$\tau$ are purely imaginary, and thus the above formula for~$R$ becomes
\[ R \;=\; R_0 \:\zeta^2\; \frac{\tau^2-1}{\tau^2 (R_0^2-|m_0|^2) - 2 \tau \:\zeta
\, {\mbox{Re}}(m_0) - \zeta^2} \:. \]
For a stationary circle, $R$ must equal~$R_0$, for all imaginary~$\tau$ (because
the range of the function~$\tanh(\zeta x)$ is $i \R$). Multiplying by the denominator, we get a quadratic
polynomial in~$\tau$, and comparing the coefficients, we obtain the conditions
\[ {\mbox{Re}}\, m_0 \;=\; 0 \qquad {\mbox{and}} \qquad
|m_0|^2 - R_0^2 \;=\; |\zeta|^2\:. \]
We thus obtain for any~$R_0 \geq 0$ the two circles with centers
\beq \label{fixedcircle}
m_0 \;=\; \pm i \sqrt{R_0^2 + |\zeta|^2} \:,
\eeq
and an easy computation shows that these circles are indeed stationary under the flow.
Since these circles enclose the fixed points (see Figure~\ref{fig0} (right)),
both fixed points are centers.

This simple analysis gives rise to a useful concept. Namely, 
the above flow of circles under the Riccati equation defines
an invariant region consisting of the interior of the circles together with their boundary.
In the next section we shall implement this point of view in the case of a
non-constant potential~$V$.

\section{Invariant Disk Estimates for the Riccati Equation}
\setcounter{equation}{0}
We now turn our attention to solutions~$y \in \C$ of the Riccati equation~\eqref{riccati}
for a general complex-valued potential~$V(x)$. Our goal is to derive inequalities of the form
\[ |y(x) - m(x)| \;\leq\; R(x) \]
for explicit functions~$m$ and~$R$. Geometrically, this means that the exact solution~$y$
stays inside the disk of radius~$R$ centered at~$m$. Thus we can interpret~$m$ as an
approximate solution, with~$R$ giving a rigorous bound on the error.

In preparation, we show that the the flow given by the Riccati equation maps circles
into circles. At the same time, we derive differential equations describing the motion of
these circles. Let us assume that~$y$ lies on a circle of radius~$R$ centered at~$m$, i.e.
\[ y \;=\; m+R \, e^{i \varphi}\:, \]
where~$R \geq 0$, $\varphi \in \R$, $m \in \C$.
We now derive differential equations for~$R$ and~$m$. First,
\begin{eqnarray*}
m'+R'e^{i \varphi} + i \varphi'\: R e^{i \varphi} &=& y' \\
&=& V - y^2 \;=\; V - m^2 - 2 R m e^{i \varphi} - R^2 e^{2 i \varphi} \:.
\end{eqnarray*}
We multiply by~$e^{-i \varphi}$ and take the real parts to get
\[ {\mbox{Re}} \left[ e^{-i \varphi} (V - m^2-m') - 2 R m - R' - i \varphi'\: R\: - e^{i \varphi}\, R^2 
\right] \;=\;0 \:. \]
In taking the real part, the term involving~$\varphi'$ drops out. Furthermore,
we may replace the factor~$e^{i \varphi}$ by~$e^{-i \varphi}$. This gives
\[ {\mbox{Re}} \left[ e^{-i \varphi} (V - m^2-m'-R^2) - 2 R m - R' \right] \;=\;0 \:. \]
Since this equation should hold for any~$\varphi$, we obtain the system of ODEs
\begin{align*}
R' &= -2 R \; {\mbox{\rm{Re}}}\, m \\
m' &= V - m^2 - R^2 \:.
\end{align*}
It is difficult to solve this system exactly. But the following lemma gives us a method to
obtain invariant disk estimates.
\begin{Lemma} \label{lemma1} For a given complex-valued potential~$V=V(x)$, we
assume that the complex function~$m$ and the real function~$R \geq 0$ satisfy
on the interval~$I := [x_0, x_1]$ the differential equations
\begin{eqnarray}
R' &=& -2 R \; {\mbox{\rm{Re}}}\, m \:+\: \delta R \label{Req} \\
m' &=& V - m^2 - R^2 \:+\: \delta m\:, \label{meq}
\end{eqnarray}
where the real function~$\delta R$ and the complex function~$\delta m$
obey the inequality
\beq \label{deltain}
\delta R \;\geq\; |\delta m|\:.
\eeq
Then the disk with radius~$R$ centered at~$m$ is invariant, i.e.\
for any solution of the Riccati equation~\eqref{riccati} the following
implication holds:
\[ |y-m|(x_0) \leq R(x_0) \quad \Longrightarrow \quad |y-m|(x) \leq R(x)\;\; \forall x \in I\:. \]
\end{Lemma}
\Proof We first prove the lemma in the case of strict inequality in~\eqref{deltain}.
If the lemma were false, there would be an $x \in I$ such that~$|y-m|(x)>R(x)$.
Let~$\bar{x} \in [x_0, x_1)$ be the infimum of all such~$x$. Then~$|y-m|(\bar{x})=R(\bar{x})$.
Using~\eqref{riccati} and~\eqref{meq}, we find at~$\bar{x}$
\begin{eqnarray*}
\frac{d}{dx} |y-m| &=& {\mbox{Re}}\: \frac{(y'-m') \overline{(y-m)}}{|y-m|}
\;=\; {\mbox{Re}}\: \frac{(V-y^2 - V + m^2 + R^2 - \delta m) \overline{(y-m)}}{|y-m|} \\
&\leq& R \:{\mbox{Re}}(y-m) +|\delta m| \:-\:{\mbox{Re}} \:\frac{(y^2-m^2) \overline{(y-m)}}{|y-m|} \\
&=& R \:{\mbox{Re}}(y-m) +|\delta m| \:-\:R\: {\mbox{Re}} (y+m) \;=\;
-2 R \; {\mbox{\rm{Re}}}\, m +|\delta m|\:.
\end{eqnarray*}
Combining this with~\eqref{Req} and the strict inequality in~\eqref{deltain}, we obtain
\[ \frac{d}{dx} |y-m|(\bar{x}) \;<\; \frac{d}{dx} R(\bar{x})\:. \]
Hence there is a~$\delta>0$ such that on the interval~$(\bar{x}, \bar{x}+\delta)$
the inequality~$|y-m|<R$ holds. This is a contradiction.

To prove the general case of non-strict inequality in~\eqref{deltain}, we introduce the new
radius~$\tilde{R}$ by
\[ \tilde{R} \;=\; R + \varepsilon e^{L (x-x_1)}\:, \]
where~$\varepsilon > 0$ and
\beq \label{Ldef}
L \;>\; \max_I \left( -2 \,{\mbox{Re}}\, m + 2 R - \varepsilon e^{L(x-x_1)} \right) .
\eeq
Then~$\tilde{R}$ and~$m$ satisfy again the differential equations~\eqref{Req} and~\eqref{meq},
where~$R$ is replaced by~$\tilde{R}$, and~$\delta R$ and~$\delta m$ are replaced by
\begin{eqnarray*}
\delta \tilde{R} &=& \delta R + \varepsilon e^{L(x-x_1)} \left( L + 2 {\mbox{Re}}\, m \right) \\
\delta \tilde{m} &=& \delta m +  \varepsilon e^{L(x-x_1)} \left( 2 R - \varepsilon e^{L(x-x_1)} \right) .
\end{eqnarray*}
Using~\eqref{deltain} and~\eqref{Ldef}, we see that~$|\delta \tilde{m}|<\delta \tilde{R}$ for
sufficiently small~$\varepsilon$,
and thus the circle with radius~$\tilde{R}$ and center~$m$ is invariant.
Letting~$\varepsilon \searrow 0$ completes the proof.
\QED

In what follows, it is convenient to write~$m=\alpha+i \beta$ with real~$\alpha$ and~$\beta$, 
and similarly we write $\delta m = \delta \alpha + i \,\delta \beta$.
Then the equations~\eqref{Req} and~\eqref{meq} become
\begin{eqnarray}
R' &=& -2 \alpha R \:+\: \delta R \label{Req2} \\
\alpha' &=& {\mbox{Re}}\, V - \alpha^2 + \beta^2 - R^2 \:+\: \delta \alpha \label{alphaeq} \\
\beta' &=& {\mbox{Im}}\, V - 2 \alpha \beta \:+\: \delta \beta\:. \label{betaeq}
\end{eqnarray}
In our first theorem we satisfy the conditions of Lemma~\ref{lemma1}
by employing a special ansatz where the invariant disks are centered on the real axis
(so that~$\beta \equiv 0$). We introduce the abbreviation
\beq
\boxed{ \quad U \;=\; {\mbox{Re}}\, V - \alpha^2 - \alpha' \:. \quad } \label{Udef}
\eeq
\begin{Thm} \label{thm2}
Suppose that on an interval~$I \subset \R$, we are given the real functions~$\alpha, W \in C^1(I)$.
Assume furthermore that~$W > 0$ and that
\beq \label{invcond}
2 \alpha W + \frac{W'}{2} - \sqrt{W}\:\Big(  |W-U| + |{\mbox{\rm{Im}}}\, V| \Big) \;\geq\; 0\:.
\eeq
Then the circle with radius~$R$ centered at~$m$ with
\beq \label{ansatz}
R \;=\; \sqrt{W}\:, \qquad m \;=\; \alpha
\eeq
is invariant on~$I$ under the Riccati flow~\eqref{riccati}.
\end{Thm}
\Proof We shall verify the conditions of Lemma~\ref{lemma1}.
Using the ansatz~\eqref{ansatz} in the equations~\eqref{Req2}, \eqref{alphaeq} and~\eqref{betaeq},
we get
\[ \delta R \;=\; \frac{W'}{2 \sqrt{W}} + 2 \alpha \sqrt{W} \:, \qquad
\delta \alpha \;=\; W-U\:,\qquad \delta \beta \;=\; -{\mbox{Im}}\, V\:. \]
The sufficient condition~$\delta R \geq |\delta \alpha|+|\delta \beta|$ reduces to~\eqref{invcond}.
\QED

In our next theorem we will not prescribe~$\beta$.
For convenience, we now replace~\eqref{deltain} by the sufficient condition
\beq \label{Rdef}
\delta R \;=\; |\delta \alpha| + |\delta \beta|\:.
\eeq
Thus our task is to find six functions~$R$, $\alpha$, $\beta$, $\delta R$, $\delta \alpha$
and~$\delta \beta$ which satisfy the equations~\eqref{Req2}, \eqref{alphaeq} and~\eqref{betaeq}
together with the condition~\eqref{Rdef}. Thus we expect that we can freely
assign two functions, and then the remaining four functions will be determined.
To explain which functions are to be preassigned, we note that the equations~\eqref{Req2}
and~\eqref{betaeq} are both linear, so that the
only nonlinearity appears in the $\alpha$-equation~\eqref{alphaeq}.
Thus we can get rid of the nonlinear differential equation by taking the functions
\begin{equation}
\alpha \qquad {\mbox{and}} \qquad W \;:=\; R^2 - \beta^2 \label{Wdef}
\end{equation}
as a-priori given. Then~\eqref{alphaeq} becomes the defining equation for~$\delta \alpha$; namely,
\[ \delta \alpha \;=\; -{\mbox{Re}}\, V + \alpha^2 + \alpha' \:+\: W\:. \]
Using these definitions, we are left with the system of equations
\begin{eqnarray}
R' &=& -2 \alpha R \:+\: |\delta \alpha|+|\delta \beta| \label{nReq} \\
\beta' &=& {\mbox{Im}}\, V - 2 \alpha \beta \:+\: \delta \beta \label{nbeq} \\
R^2 - \beta^2 &=& W \label{algebra}
\end{eqnarray}
for the three unknown functions~$R$, $\beta$ and~$\delta \beta$. This system consists of
two linear differential equations and one nonlinear algebraic equation. Of course, we must
also ensure that $0 \leq R < \infty$.

To analyze these equations, we consider the two cases~$\delta \beta \geq 0$ and~$\delta \beta<0$ separately.
\begin{description}
\item[Case (A) $\delta \beta \geq 0$]
Subtracting~\eqref{nbeq} from~\eqref{nReq} we get
\beq \label{leq1}
(R-\beta)' \;=\; -2 \alpha\, (R-\beta) - {\mbox{\rm{Im}}}\, V + |\delta \alpha|\:.
\eeq
Furthermore, differentiating~\eqref{algebra} gives
\beq \label{leq2}
2 R R' - 2 \beta \beta' \;=\; W' \:.
\eeq
We consider \eqref{leq1} and~\eqref{leq2} as a system of two linear algebraic equations
in the unknowns~$R'$ and~$\beta'$.
Solving for~$\beta'$ and substituting into~\eqref{nbeq}, we find
\beq \label{betaform}
(R-\beta)\: \delta \beta \;=\; -R\, |\delta \alpha| + 2 W \alpha + {\mbox{\rm{Im}}} V\: \beta
+ \frac{W'}{2} \:.
\eeq
Since we assumed~$\delta \beta$ to be positive, we get the consistency condition
\beq \label{concon1}
(R-\beta) \left[  2 \alpha W + \frac{W'}{2} -R\, |\delta \alpha| + \beta\: {\mbox{\rm{Im}}} V \right]  \;\geq\; 0\:.
\eeq
 Provided that this consistency condition holds, we can use~\eqref{leq1} to solve
 for~$R-\beta$. Then using~\eqref{algebra}, we conclude that~$R+\beta = W/(R-\beta)$.
\item[Case (B) $\delta \beta < 0$]
Adding~\eqref{nReq} and~\eqref{nbeq} gives
\beq \label{leq3}
(R+\beta)' \;=\; -2 \alpha\, (R+\beta) + {\mbox{\rm{Im}}}\, V + |\delta \alpha|\:.
\eeq
Solving~\eqref{leq3} and~\eqref{leq2} for~$\beta'$ and substituting the result into~\eqref{nbeq},
we obtain
\[ (R+\beta)\: \delta \beta \;=\; R\, |\delta \alpha| - 2 W \alpha - {\mbox{\rm{Im}}} V\: \beta
- \frac{W'}{2} \:, \]
giving rise in this case to the consistency condition
\beq \label{concon2}
(R+\beta) \left[  2 \alpha W + \frac{W'}{2} -R\, |\delta \alpha| + \beta\: {\mbox{\rm{Im}}} V \right] \;\geq\; 0\:.
\eeq
 Provided that this consistency condition holds, $R+\beta$ is determined by~\eqref{leq3},
 and $R-\beta = W/(R+\beta)$.
\end{description}

Since the bracket in~\eqref{concon1} and~\eqref{concon2} will appear often, it is convenient to
introduce the abbreviation
\begin{equation}
\boxed{ \quad
{\mathfrak{D}} =
 2 \alpha W + \frac{W'}{2} -R\, |\delta \alpha| + \beta\: {\mbox{\rm{Im}}} V \: . \quad }
\label{determ}
\end{equation}
We refer to~${\mathfrak{D}}$ as the~{\bf{determinator}}.
Then the following invariant disk estimate holds.
\begin{Thm} \label{thm1}
Suppose that on an interval~$I \subset \R$, we are given the real functions~$\alpha, W \in C^1(I)$.
We set
\begin{equation} \label{sigmadef}
\sigma(x) \;=\; \exp \left( \int^x 2 \alpha \right) .
\end{equation}
Then the following statements hold.
\begin{itemize}
\item[{\bf{(A)}}] Define real functions~$R$ and~$\beta$ on~$I$ by
\begin{eqnarray}
(R-\beta)(x) &=& \frac{1}{\sigma} \int^x \sigma
\left(  - {\mbox{\rm{Im}}}\, V + |W-U| \right) \label{Rmb} \\[.5em] 
(R+\beta)(x) &=& \frac{W(x)}{(R-\beta)(x)} \label{Rpb}
\end{eqnarray}
(with~$U$ as defined by~\eqref{Udef}).
Assume that the function~$R-\beta$ has no zeros, $R \geq 0$, and that
\beq \label{bc1}
(R-\beta) \:{\mathfrak{D}} \;\geq\; 0
\eeq
(with~${\mathfrak{D}}$ according to~\eqref{determ}).
Then the circle centered at~$m(x)=\alpha+i \beta$ with radius~$R(x)$ is invariant 
on~$I$ under the Riccati flow~\eqref{riccati}.
\item[{\bf{(B)}}] Define real functions~$R$ and~$\beta$ on~$I$ by
\begin{eqnarray}
(R+\beta)(x) &=& \frac{1}{\sigma} \int^x \sigma
\left(  {\mbox{\rm{Im}}}\, V + |W-U| \right) \label{Rmb2} \\[.5em]
(R-\beta)(x) &=& \frac{W(x)}{(R+\beta)(x)}\:. \label{Rpb2}
\end{eqnarray}
Assume that the function~$R+\beta$ has no zeros, $R \geq 0$, and that
\beq \label{condition2}
(R+\beta) \: {\mathfrak{D}} \;\geq\; 0 \:.
\eeq
Then the circle centered at~$m(x)=\alpha+i \beta$ with radius~$R(x)$ is invariant on~$I$
under the Riccati flow~\eqref{riccati}.
\end{itemize}
\end{Thm}
\Proof We only prove the first part, as the proof of the second part is similar.
Since~\eqref{bc1} holds, we are consistently in Case~(A) above.
Integrating the differential equation~\eqref{leq1} gives~\eqref{Rmb}, whereas~\eqref{Rpb}
follows from~\eqref{Wdef}. We thus have a solution of the system of
equations~\eqref{nReq}, \eqref{nbeq} and~\eqref{algebra} with~$\delta \beta$ given by~\eqref{betaform}.
Applying Lemma~\ref{lemma1} completes the proof.
\QED

We next outline the general strategy for applying the last theorem. Suppose that
an approximate solution~$\tilde{y}$ of the Riccati equation is given.
Setting~$\tilde{V} = \tilde{y}' + \tilde{y}^2$, we can consider~$\tilde{y}$ as a solution
of a Riccati equation
\beq \label{approxric}
\tilde{y}' \;=\; \tilde{V} - \tilde{y}^2
\eeq
with an ``approximate'' potential $\tilde{V}$. Since~$\tilde{y}$ approximates~$y$,
an obvious idea would be to consider circles centered at~$m=\tilde{y}$.
In the estimates of Theorem~\ref{thm1}, however, we only need to prescribe the function
$\alpha={\mbox{Re}}\, m$, whereas the imaginary part of~$m$ will then be determined.
Therefore, it is natural to choose
\beq \label{alphachoice}
\alpha \;=\; {\mbox{Re}}\, \tilde{y}\:.
\eeq
A natural choice for~$W$ is to set
\beq \label{Wchoice}
W \;=\; U
\eeq
(with~$U$ according to~\eqref{Udef}), because then the term $|W-U|$ in~\eqref{Rmb} and~\eqref{Rmb2} vanishes\footnote{Other choices of~$W$ might be useful, but we will not consider
them in this paper.}. Then Theorem~\ref{thm1} gives us invariant disk estimates.
The center~$m$ of this disk can be viewed as an improved approximate solution, and its error,
measured as the distance from the exact solution, is at most~$R$. The invariant disk estimate depends crucially
on the sign of the determinator~${\mathfrak{D}}$.
Thus in applications it is important to know the sign of~${\mathfrak{D}}$.
It can even be useful to modify~$\tilde{y}$ so as to give~${\mathfrak{D}}$
a desired sign (this will be illustrated in Section~\ref{sec5} in a few examples).
For this purpose, it is useful to bring the determinator into a more convenient form.
\begin{Lemma} \label{lemma34}
Suppose that, for a given solution~$\tilde{y}$ of an approximate
Riccati equation~\eqref{approxric}, $\alpha$ and~$W$ are chosen according to~\eqref{alphachoice}
and~\eqref{Wchoice}. Then the determinator is given by
\beq \label{Dform}
{\mathfrak{D}} \;=\; 2 \alpha\, {\mbox{\rm{Re}}} (V-\tilde{V}) \:+\:
\frac{1}{2}\: {\mbox{\rm{Re}}} (V-\tilde{V})' - \tilde{\beta}\: {\mbox{\rm{Im}}} \,\tilde{V}
+ \beta\: {\mbox{\rm{Im}}} \,V \:,
\eeq
where~$\tilde{\beta}={\mbox{\rm{Im}}}\, \tilde{y}$.
\end{Lemma}
\Proof Writing the Riccati equation~\eqref{approxric} as
\[ \alpha' \;=\; {\mbox{Re}}\, \tilde{V} - \alpha^2 + \tilde{\beta}^2\:,\qquad
\tilde{\beta}' \;=\; {\mbox{Im}}\, \tilde{V} - 2 \alpha \tilde{\beta}\:, \]
we can use~\eqref{Wchoice} and~\eqref{Udef} to obtain
\[ W \;=\; U \;=\; {\mbox{Re}} (V-\tilde{V}) - \tilde{\beta}^2\:. \]
Applying these identities in~\eqref{determ} gives
\[ {\mathfrak{D}} \;=\; 2 \alpha\, {\mbox{\rm{Re}}} (V-\tilde{V}) \:+\:
\frac{1}{2}\: {\mbox{\rm{Re}}} (V-\tilde{V})' - 2 \alpha \tilde{\beta}^2 - \tilde{\beta}
\tilde{\beta}' + \beta\: {\mbox{\rm{Im}}} \,V\:. \]
Using the~$\tilde{\beta}'$-equation finishes the proof.
\QED

\section{Error Estimates for Real Potentials and Examples}
\setcounter{equation}{0}
In this section we shall illustrate our error estimates in examples for a real potential~$V$ and explain
how the previous results from~\cite{FS, FKSY2} fit into our framework.
We first mention an additional structural relation which appears only for a real potential.
Namely, suppose that~$\phi$ is a complex solution of the Schr\"odinger
equation~\eqref{schrodinger}. Decomposing it into its real and imaginary parts,
$\phi = \phi_1 + i \phi_2$, the functions~$\phi_1$ and~$\phi_2$ are real solutions of the
Schr\"odinger equation forming a fundamental set. Thus their Wronskian
$w = \phi_1 \phi_2' - \phi_1' \phi_2$ is a constant. It can be expressed as
$w = {\mbox{Im}}\: (\overline{\phi}\: \phi' )
= |\phi|^2\: {\mbox{Im}}\:y$, giving rise to the relation
\[  |\phi|^2 \;=\; \frac{w}{\mbox{Im}\:y}\;, \]
which is useful for estimating the amplitude of~$\phi$. Moreover, this relation
shows that the imaginary part of~$y$ can never change sign,
and thus the upper and lower half planes must be invariant regions for~$y$.

\begin{figure}[tbp]%
\begin{center}%
\begin{picture}(0,0)%
\includegraphics{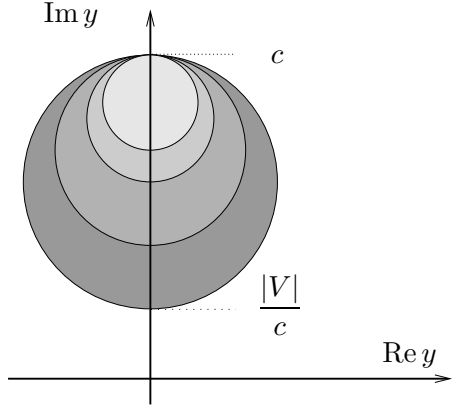}%
\end{picture}%
\setlength{\unitlength}{1575sp}%
\begingroup\makeatletter\ifx\SetFigFont\undefined%
\gdef\SetFigFont#1#2#3#4#5{%
  \reset@font\fontsize{#1}{#2pt}%
  \fontfamily{#3}\fontseries{#4}\fontshape{#5}%
  \selectfont}%
\fi\endgroup%
\begin{picture}(7041,6444)(1228,-7699)
\put(5399,-2296){\makebox(0,0)[lb]{\smash{{{$c$}%
}}}}
\put(1824,-1642){\makebox(0,0)[lb]{\smash{{${\mbox{Im}}\, y$}}}}
\put(7156,-6991){\makebox(0,0)[lb]{\smash{{${\mbox{Re}}\, y$}}}}
\put(5179,-6265){\makebox(0,0)[lb]{\smash{{{$\displaystyle \frac{|V|}{c}$}%
}}}}
\end{picture}%
\caption{A simple invariant disk estimate for a negative, increasing potential.}%
\label{fig1}%
\end{center}%
\end{figure}

To illustrate how our estimates apply, we begin with a simple example.
\begin{Example} {\bf{(negative, increasing potential)}} {\em{Suppose that the
potential satisfies on the interval~$I=[x_0, x_1]$ the conditions
\[ V(x) \;\leq\; 0  \:,\qquad V'(x) \;\geq\; 0\:. \]
The simplest (but certainly not the best) choice is to set~$\alpha \equiv 0$.
Furthermore, setting~$W=U$, we see from~\eqref{sigmadef} that~$\sigma$ is a constant.
According to~\eqref{Udef}, we find that~$W=V$. The determinator~\eqref{determ}
simplifies to
\beq \label{Dex}
{\mathfrak{D}} \;=\; \frac{V'}{2} \;\geq\; 0 \:.
\eeq
We want to apply part~(B) of Theorem~\ref{thm1} (part~(A) is similar
if one flips the sign of~$\beta$). From~\eqref{Rmb2} we see that~$R+\beta$ is a constant,
which according to~\eqref{condition2} and~\eqref{Dex} must be positive. We thus obtain
\beq \label{br}
\beta+R \;=\; c >0 \:, \qquad \beta-R \;=\; \frac{|V|}{c}\:.
\eeq
The condition~$R \geq 0$ gives the constraint
\[ c^2 \;\geq\; \max_I |V| \;=\; |V(x_0)|\:. \]
If the potential is constant, we recover the stationary circles of Section~\ref{sec2}
(as is easily verified by comparing~\eqref{br} and~\eqref{fixedcircle}).
If the potential is non-constant, we
get an increasing family of invariant disks in the upper half plane centered
on the imaginary axis whose highest point
is fixed at $i c$, and whose lowest point varies like~$i |V|/c$; see Figure~\ref{fig1}.
\QEDrem
}}\end{Example}

\begin{figure}[tbp]%
\begin{center}%
\begin{picture}(0,0)%
\includegraphics{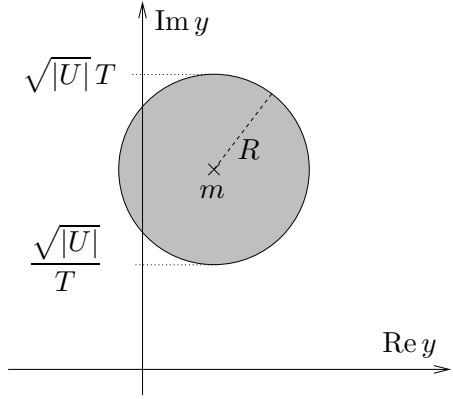}%
\end{picture}%
\setlength{\unitlength}{1575sp}%
\begingroup\makeatletter\ifx\SetFigFont\undefined%
\gdef\SetFigFont#1#2#3#4#5{%
  \reset@font\fontsize{#1}{#2pt}%
  \fontfamily{#3}\fontseries{#4}\fontshape{#5}%
  \selectfont}%
\fi\endgroup%
\begin{picture}(7019,6254)(1239,-7688)
\put(4876,-3946){\makebox(0,0)[lb]{\smash{{\SetFigFont{11}{13.2}{\familydefault}$R$}}}}
\put(3571,-1921){\makebox(0,0)[lb]{\smash{{\SetFigFont{11}{13.2}{\familydefault}${\mbox{Im}}\, y$}}}}
\put(7156,-6991){\makebox(0,0)[lb]{\smash{{\SetFigFont{11}{13.2}{\familydefault}${\mbox{Re}}\, y$}}}}
\put(4276,-4561){\makebox(0,0)[lb]{\smash{{\SetFigFont{11}{13.2}{\familydefault}$m$}}}}
\put(1576,-5716){\makebox(0,0)[lb]{\smash{{\SetFigFont{11}{13.2}{\familydefault}$\displaystyle \frac{\sqrt{|U|}}{T}$}}}}
\put(1486,-2761){\makebox(0,0)[lb]{\smash{{\SetFigFont{11}{13.2}{\familydefault}$\sqrt{|U|}\, T$}}}}
\end{picture}%
\caption{Invariant disk estimate for~$U<0$.}%
\label{fig2}%
\end{center}%
\end{figure}

We next consider the situation for a general real potential~$V$ and a general choice
of~$\alpha$, again setting~$W=U$. Then the determinator becomes
\beq \label{mkDeq}
\mathfrak{D} \;=\; 2 \alpha U + \frac{U'}{2} \;=\; \frac{U}{2}\:
\frac{(\sigma^2 U)'}{\sigma^2 U}
\eeq
(with~$\sigma$ defined by~\eqref{sigmadef}). The main simplification is that
now~$\mathfrak{D}$ is an a-priori given function. In the case when~$R-\beta$ and~$R+\beta$
have opposite signs, we see from~\eqref{bc1} and~\eqref{condition2} that, depending on the
sign of~$\mathfrak{D}$, either part~(A) or part~(B) of Theorem~\ref{thm1} applies.
In this case, we can combine parts~(A) and~(B) to obtain a statement which can be used even
if~$\mathfrak{D}$ changes sign infinitely often. In view of~\eqref{algebra},
the condition that~$R-\beta$ and~$R+\beta$ should have opposite signs is equivalent to
the simpler condition that~$U<0$; it means that the invariant disk either lies in the upper
half plane or in the lower half plane, without intersecting the real line. By continuity, the invariant disk cannot move from the upper to the lower half plane. Thus we can assume without loss of generality that the invariant disk lies in the upper half plane. This means that~$R+\beta >0$
and~$R-\beta<0$. It is convenient to satisfy the equation~\eqref{algebra}
(and thus also~\eqref{Rpb} and~\eqref{Rpb2}) by making the ansatz
\beq \label{bransatz}
\beta \;=\; \frac{\sqrt{|U|}}{2} \left(T + \frac{1}{T} \right) , \spc
R \;=\; \frac{\sqrt{|U|}}{2} \left(T - \frac{1}{T} \right)
\eeq
with a free function~$T>1$. If~$\mathfrak{D}$ is negative, we apply Theorem~\ref{thm1}~(A)
to obtain
\[ T \;=\; -\frac{\sqrt{|U|}}{R-\beta} \;=\; -\frac{\sigma \sqrt{|U|}}{c} \]
with~$c$ an integration constant. Diffentiating gives
\[ \frac{T'}{T} \;=\; \frac{(\sigma \sqrt{|U|})'}{\sigma \sqrt{|U|}} \;=\;
\frac{1}{2}\: \frac{(\sigma^2 |U|)'}{\sigma^2 |U|} \;=\;
\frac{1}{2}\: \frac{(\sigma^2 U)'}{\sigma^2 U} 
\;=\; \frac{1}{2}\: \left| \frac{(\sigma^2 U)'}{\sigma^2 U} \right| , \]
where in the last two steps we used that~$U$ is negative, and that in the considered
case~$\mathfrak{D}<0$, the term~$(\sigma^2 U)'/(\sigma^2 U)$ is positive according
to~\eqref{mkDeq}.
If~$\mathfrak{D}$ is positive, we apply Theorem~\ref{thm1}~(B) to get
\[ T \;=\; \frac{R+\beta}{\sqrt{|U|}} \;=\; \frac{c}{\sigma \sqrt{|U|}} \:. \]
Now differentiating gives
\[ \frac{T'}{T} \;=\; -\frac{(\sigma \sqrt{|U|})'}{\sigma \sqrt{|U|}} \;=\;
-\frac{1}{2}\: \frac{(\sigma^2 U)'}{\sigma^2 U} 
\;=\; \frac{1}{2}\: \left| \frac{(\sigma^2 U)'}{\sigma^2 U} \right| , \]
where in the last step we used that now~$(\sigma^2 U)'/(\sigma^2 U)<0$.
We conclude that, independent of the sign of~$\mathfrak{D}$, we obtain the
differential equation
\[ \frac{T'}{T} \;=\;  \frac{1}{2}\: \left| \frac{(\sigma^2 U)'}{\sigma^2 U} \right| . \]
Integrating both sides, we can rewrite~$T$ as a total variation,
\beq \label{Tdef}
T(x) \;=\; T_0\: \exp \left( \frac{1}{2}\: {\mbox{TV}}_{[x_0,x)}
\log |\sigma^2 U| \right) .
\eeq
In this way, we recover the invariant disk estimate which was first obtained
in~\cite[Lemma~4.1]{FS}; see Figure~\ref{fig2}.
\begin{Lemma} \label{lemmainv}
Let~$\alpha$ be a real function on $[x_0,x_1]$ which is continuous and
piecewise $C^1$.
Suppose that the function $U := V-\alpha^2-\alpha'$ is negative on~$[x_0,x_1]$.
Using the above definitions~\eqref{Tdef} and~\eqref{bransatz}, the disks
of radius~$R$ centered at~$m=\alpha + i \beta$ are invariant under the Riccati
flow~\eqref{riccati}, on~$[x_0,x_1]$.
\end{Lemma}
We now illustrate in two examples how this lemma can be applied (for other applications
see~\cite[Theorems~4.2 and~4.3]{FS} and~\cite[Lemmas~4.10 and~4.12]{FKSY2}).
In the first example, we consider the case that the potential~$V$ is negative.
We recall that the WKB wave functions are given by (see for example~\cite[Section~2.4]{Sakurai}
or~\cite{FF})
\beq \label{WKBwaves}
\phi_{\mbox{\tiny{WKB}}}(x) \;=\; |V(x)|^{-\frac{1}{4}}\:
\exp \left( \pm i \int^x \sqrt{|V|} \right) .
\eeq
A short calculation shows that these functions satisfy the equation
\[ \frac{d^2}{dx^2} \phi_{\mbox{\tiny{WKB}}} - V \phi_{\mbox{\tiny{WKB}}} =
\left( \frac{5}{16}\: \frac{|V|'^2}{|V|^2} - \frac{1}{4}\: \frac{|V|''}{|V|} \right) 
\phi_{\mbox{\tiny{WKB}}} \:. \]
Interpreting the right side as an error term, one sees that the
WKB wave functions are expected to be a good approximation to~$\phi$ provided that
\beq \label{WKBcond}
\left| \frac{V''}{V^2} \right| + \left| \frac{V'^2}{V^3} \right| \;\ll\; 1\:.
\eeq
For definiteness, we consider the plus sign in~\eqref{WKBwaves} (the minus sign
can be obtained by complex conjugation). Then the corresponding approximate
Riccati solution is given by
\[ y_{\mbox{\tiny{WKB}}} \;=\; \frac{\phi'_{\mbox{\tiny{WKB}}}}{\phi_{\mbox{\tiny{WKB}}}} \;=\; i \sqrt{|V|} - \frac{V'}{4 V} \:. \]
Following the general strategy described after Theorem~\ref{thm1}, we thus choose
\beq \label{aldef}
\alpha \;=\; {\mbox{Re}}\, y_{\mbox{\tiny{WKB}}} \;=\;
 - \frac{V'}{4 V}\:.
 \eeq
An easy computation gives
\begin{eqnarray}
\alpha' &=& -\frac{V''}{4V} + \frac{V'^2}{4 \:V^2} \\
U &=& V \left(1 + \frac{V''}{4 V^2} \:+\: \frac{5\:V'^2}{16\:|V|^3}
\right) \label{UWKB} \\
\sigma &=& \exp \left(-\frac{1}{2} \int^x \frac{V'}{V} \right) \;=\; \frac{c}{\sqrt{|V|}} \\
\sigma^2 U &=& -c^2 \left(1 + \frac{V''}{4 V^2} \:+\: \frac{5\:V'^2}{16\:|V|^3}
\right) .
\end{eqnarray}
In view of~\eqref{UWKB}, the condition~\eqref{WKBcond} and the fact
that~$V<0$, we see that~$U$ is negative.
Thus Lemma~\ref{lemmainv} applies, giving the following estimate.
\begin{Example} {\bf{(error estimate for WKB wave function, $V<0$)}} \\
Suppose that~$V$ is negative and
\[ -\frac{V''}{4 V^2} \:-\: \frac{5\:V'^2}{16\:|V|^3} \;<\;1\:. \]
Then the disks of radius~$R$ centered at~$m=\alpha + i \beta$ as given by~\eqref{aldef},
\eqref{bransatz} and
\beq \label{Trel2}
T(x) \;=\; T_0 \: \exp \left( \frac{1}{2}\: {\mbox{TV}}_{[x_0,x)}
\log \left[ 1 + \frac{V''}{4 V^2} \:+\: \frac{5\:V'^2}{16\:|V|^3} \right] \right)
\eeq
are invariant under the Riccati flow~\eqref{riccati}, on~$[x_0,x_1]$.
\end{Example}
In our setting, the semiclassical limit can be described by scaling the potential
according to~$V \rightarrow \lambda V$ and considering the behavior as~$\lambda \rightarrow \infty$.
In this limit, the square bracket in~\eqref{Trel2} converges to one, so that the logarithm
vanishes. As a consequence, T becomes a constant. Choosing~$T_0=1$, we find that the radius~$R(x)$ as
given by~\eqref{bransatz} vanishes identically in the semiclassical limit, in agreement with the
fact that the WKB wave function goes over to the exact solution.

The next estimate even applies near a zero of the potential.
\begin{Example} {\bf{(invariant disk estimate with exponential bound)}} \\
Choose a constant~$\alpha$,
\beq \label{aldef2}
\alpha \;=\; c+ \sup_{x \in [x_0, x_1]} \sqrt{ \max(0, V(x)) }
\eeq
with~$c \geq 0$. Then
\begin{eqnarray*}
U &=& V-\alpha^2 \;<\; -c^2 \;\leq\; 0 \\
\sigma &=& C\, e^{2 \alpha x} \\
\sigma^2 U &=&  C^2\, e^{4 \alpha x} \left(V-\alpha^2 \right)
\end{eqnarray*}
with an integration constant~$C$.
Hence Lemma~\ref{lemmainv} yields that the disks of radius~$R$ centered at~$m=\alpha + i \beta$ as
given by~\eqref{aldef2}, \eqref{bransatz} and
\[ T(x) \;=\; T_0 \: \exp \left( \frac{1}{2}\: {\mbox{TV}}_{[x_0,x)}
\log \left[ e^{4 \alpha x} \left(V-\alpha^2 \right) \right] \right) \]
are invariant under the Riccati flow~\eqref{riccati}, on~$[x_0,x_1]$.
\end{Example}
Clearly, due to the exponential factor~$e^{4 \alpha x}$, this estimate gets weaker as the size
of the interval increases. We shall below see a much better estimate using Airy functions
(see Example~\ref{ex1}).

We next consider how Theorem~\ref{thm1} applies in the case~$U \geq 0$
(where we again choose~$W=U$). Then, according to~\eqref{algebra},
the functions~$R-\beta$ and~$R+\beta$ have the same sign, which must be positive
because~$R \geq 0$. Inspection of~\eqref{bc1} and~\eqref{condition2} yields that
the determinator must be positive, i.e.
\beq \label{Upcond}
U' + 4 \alpha U \;\geq\; 0 \:.
\eeq
Provided that this condition holds, we may apply both parts~(A) and~(B) of Theorem~\ref{thm1}.
Taking the intersection of the corresponding invariant disks gives the following result,
which generalizes~\cite[Lemma~4.2]{FKSY2}.
\begin{Lemma} \label{lemmainv2}
Let~$\alpha$ be a real function on $[x_0,x_1]$ which is continuous and
piecewise $C^1$.
Suppose that the function $U := V-\alpha^2-\alpha'$ is positive on~$[x_0,x_1]$
and that the condition~\eqref{Upcond} holds. For any positive constants~$c_k$, $k=1,2$,
we introduce the two disks of radii~$R_k$ centered at~$m_k = \alpha + i \beta_k$
with
\begin{eqnarray*}
R_1 &=& \frac{1}{2} \left( \frac{U \sigma}{c_1} + \frac{c_1}{\sigma} \right) \:,\qquad
\beta_1 \;=\; \frac{1}{2} \left( \frac{U \sigma}{c_1} - \frac{c_1}{\sigma} \right) \\
R_2 &=& \frac{1}{2} \left( \frac{U \sigma}{c_2} + \frac{c_2}{\sigma} \right) \:,\qquad
\beta_2 \;=\; -\frac{1}{2} \left( \frac{U \sigma}{c_2} - \frac{c_2}{\sigma} \right) .
\end{eqnarray*}
Then the two disks as well as the lens-shaped region obtained as their intersection
are invariant under the Riccati flow~\eqref{riccati}.
\end{Lemma}
\begin{figure}[tbp]%
\begin{center}%
\begin{picture}(0,0)%
\includegraphics{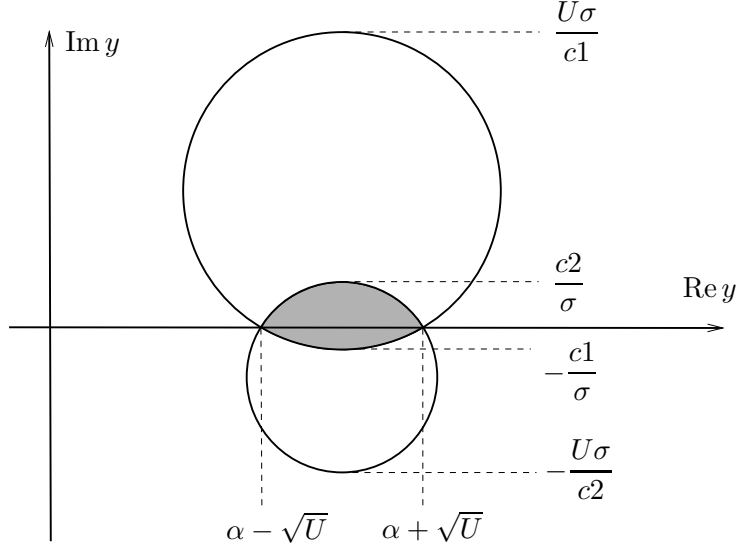}%
\end{picture}%
\setlength{\unitlength}{1575sp}%
\begingroup\makeatletter\ifx\SetFigFont\undefined%
\gdef\SetFigFont#1#2#3#4#5{%
  \reset@font\fontsize{#1}{#2pt}%
  \fontfamily{#3}\fontseries{#4}\fontshape{#5}%
  \selectfont}%
\fi\endgroup%
\begin{picture}(11318,8956)(1228,-10653)
\put(9829,-2644){\makebox(0,0)[lb]{\smash{{{$\displaystyle \frac{U \sigma}{c1}$}%
}}}}
\put(4679,-10503){\makebox(0,0)[lb]{\smash{{{$\alpha-\sqrt{U}$}%
}}}}
\put(9803,-6596){\makebox(0,0)[lb]{\smash{{{$\displaystyle \frac{c2}{\sigma}$}%
}}}}
\put(9662,-8000){\makebox(0,0)[lb]{\smash{{{$\displaystyle -\frac{c1}{\sigma}$}%
}}}}
\put(9695,-9548){\makebox(0,0)[lb]{\smash{{{$\displaystyle -\frac{U \sigma}{c2}$}%
}}}}
\put(2147,-2810){\makebox(0,0)[lb]{\smash{{${\mbox{Im}}\, y$}}}}
\put(11863,-6684){\makebox(0,0)[lb]{\smash{{${\mbox{Re}}\, y$}}}}
\put(7125,-10483){\makebox(0,0)[lb]{\smash{{{$\alpha + \sqrt{U}$}%
}}}}
\end{picture}%
\caption{Invariant lens-shaped region estimate for~$U>0$.}%
\label{fig3}%
\end{center}%
\end{figure}%
As is easily verified, the boundaries of both disks intersect the real axis at the points
$\alpha\pm \sqrt{U}$, and thus their intersection is non-void; see Figure~\ref{fig3}.

In our next example we shall apply Lemma~\ref{lemmainv2} to the WKB wave function
for a positive potential. We thus assume that~$V>0$ and that~\eqref{WKBcond} is again satisfied.
Considering the WKB wave function
\beq \label{WKB2}
\phi_{\mbox{\tiny{WKB}}}(x) \;=\; V(x)^{-\frac{1}{4}}\:
\exp \left( \int^x \sqrt{V} \right) ,
\eeq
an easy computation shows that
\begin{eqnarray*}
y_{\mbox{\tiny{WKB}}}(x) &=& \frac{\phi'_{\mbox{\tiny{WKB}}}}
{\phi_{\mbox{\tiny{WKB}}}} \;=\; \sqrt{V} - \frac{V'}{4 V} \\
\alpha &=& {\mbox{Re}}\, y_{\mbox{\tiny{WKB}}}(x) \;=\;  \sqrt{V} - \frac{V'}{4 V} \\
U &=& V - \alpha^2 - \alpha' \;=\; -\frac{5}{16}\: \frac{V'^2}{V^2} + \frac{V''}{4 V}\:.
\end{eqnarray*}
Unfortunately, in contrast to our previous example~\eqref{UWKB}, now~$U$ need not have
a fixed sign, and thus one cannot apply the above lemmas. Our method for getting around this
problem is to consider in the WKB ansatz another potential~$V_{\mbox{\tiny{WKB}}}$, and to set
\[ \phi_{\mbox{\tiny{WKB}}}(x) \;=\; V_{\mbox{\tiny{WKB}}}(x)^{-\frac{1}{4}}\:
\exp \left( \int^x \sqrt{V_{\mbox{\tiny{WKB}}}} \right) . \]
Then
\begin{eqnarray}
\alpha &=& {\mbox{Re}}\, \frac{\phi'_{\mbox{\tiny{WKB}}}}
{\phi_{\mbox{\tiny{WKB}}}} \;=\;  \sqrt{V_{\mbox{\tiny{WKB}}}} - \frac{V_{\mbox{\tiny{WKB}}}'}{4 V_{\mbox{\tiny{WKB}}}} \\
U &=& V - \alpha^2 - \alpha' \;=\; V-V_{\mbox{\tiny{WKB}}}
-\frac{5}{16}\: \frac{V_{\mbox{\tiny{WKB}}}'^2}{V_{\mbox{\tiny{WKB}}}^2} + \frac{V_{\mbox{\tiny{WKB}}}''}{4 V_{\mbox{\tiny{WKB}}}}\:.
\end{eqnarray}
By arranging that~$U$ has a definite sign, we may apply either Lemma~\ref{lemmainv}
or Lemma~\ref{lemmainv2}. To give a simple example, we choose
\[ V_{\mbox{\tiny{WKB}}} \;=\; \frac{V}{4}\:. \]
Then~$U$ is positive because~$V$ is positive and~\eqref{WKBcond} holds. Furthermore, a direct
computation gives
\begin{eqnarray}
\sigma &=& \exp \left( \int^x 2 \alpha \right) \;=\;
\frac{1}{\sqrt{V}}\: \exp \left( \int^x \sqrt{V} \right) \label{sdef} \\
\frac{2}{V}\: U &=& \frac{3}{2} +\left[ - \frac{5}{8}\: \frac{V'^2}{V^3} + \frac{V''}{2 V^2} \right] 
\label{sb1} \\
\frac{2}{3 V^\frac{3}{2}} \left(U'+4 \alpha U \right) &=&
1 + \left[ - \frac{5}{12}\: \frac{V'^2}{V^3} + \frac{5}{8}\: \frac{V'^3}{V^{\frac{9}{2}}}
+ \frac{V''}{3 V^2} - \frac{3 V' V''}{4 V^{\frac{7}{2}}} + \frac{V'''}{6 V^{\frac{5}{2}}} \right] .
\label{sb2}
\end{eqnarray}
We are now in a position to apply Theorem~\ref{thm1}.

\begin{Example} {\bf{(error estimate for WKB wave function, $V>0$)}} \\
Suppose that~$V$ is positive and that the square brackets in~\eqref{sb1} and~\eqref{sb2} are
both greater than minus one. Then choosing~$U$ and~$\sigma$ according to~\eqref{sb1}
and~\eqref{sdef}, the lens-shaped region from Lemma~\ref{lemmainv2}
(see Figure~\ref{fig3}) is invariant under the Riccati flow~\eqref{riccati}, on~$[x_0,x_1]$.
\end{Example}
It is worth noting that from~\eqref{sdef}, we see that~$\sigma$ grows exponentially.
Hence the lens-shaped region gets thinner, improving the estimate exponentially fast.

The previous estimates using Lemma~\ref{lemmainv} and Lemma~\ref{lemmainv2} have
the disadvantage that if~$U$ changes sign, different kinds of estimates must be pasted together.
This makes it necessary to match invariant disks (see~\cite[Lemma~4.4]{FKSY2}).
In the next section we explain a more convenient method in the more general context of a complex
potential (but also working for real potentials) which makes it possible for the invariant disks
to flow continuously across the real line.

\section{Error Estimates for Approximate WKB/Airy Solutions for Real or Complex Potentials} \label{sec5}
\setcounter{equation}{0}
In this section we illustrate the general technique for applying Theorem~\ref{thm2} and Theorem~\ref{thm1}
by discussing typical examples. 
Based on Theorems~\ref{thm2} and~\ref{thm1}, we will derive {\em{rigorous error estimates}}
for the standard approximate solutions obtained by glueing together WKB and Airy wave functions.

For a given potential (real or complex), one can usually distinguish regions where~$|V|$ is large, so that
the solutions of the Schr\"odinger equation are well-approximated by WKB wave functions.
In the remaining regions, one can approximate the potential by a linear potential, so that the
Schr\"odinger equation can be solved explicitly in terms of Airy functions.
To be more flexible, in the WKB region we consider the approximate solutions
\beq \label{VWKB}
\phi_{\mbox{\tiny{WKB}}}(x) \;=\; V_{\mbox{\tiny{WKB}}}(x)^{-\frac{1}{4}}\:
\exp \left( \int^x \pm \sqrt{V_{\mbox{\tiny{WKB}}}} \right)
\eeq
with a potential~$V_{\mbox{\tiny{WKB}}} \approx V$, which will be determined later.
Likewise, in the Airy region, our approximate wave functions are solutions of the Schr\"odinger equation
\beq \label{VA}
\phi_{\mbox{\tiny{A}}}''(x) \;=\; V_{\mbox{\tiny{A}}}\: \phi_{\mbox{\tiny{A}}}\:,
\eeq
where~$V_{\mbox{\tiny{A}}} \approx V$ is a linear function, also to be determined later.
By the standard $C^1$-glueing of the WKB and Airy functions, we obtain an approximate
wave function~$\tilde{\phi}$.
The wave function~$\tilde{\phi}$ is a weak solution of a Schr\"odinger equation
\beq \label{tVpot1} 
\tilde{\phi}'' \;=\; \tilde{V}\: \tilde{\phi} \:,
\eeq
where~$\tilde{V}$ in the Airy region coincides with~$V_{\mbox{\tiny{A}}}$, whereas in the
WKB region a short calculation gives
\beq \label{tVdef}
\tilde{V} \;=\; V_{\mbox{\tiny{WKB}}} +\frac{5}{16}\: \frac{V'^2_{\mbox{\tiny{WKB}}}}{V^2_{\mbox{\tiny{WKB}}}} - \frac{V''_{\mbox{\tiny{WKB}}}}{4 V_{\mbox{\tiny{WKB}}}} \:.
\eeq
Note that~$\tilde{V}$ is piecewise smooth, but in general has discontinuities.
Following the strategy explained after Theorem~\ref{thm1}, we introduce the function~$\tilde{y} = \tilde{\phi}'/\tilde{\phi}$, which is continuous and piecewise smooth, satisfying the Riccati equation
\beq \label{tVpot2}
\tilde{y}' \;=\; \tilde{V}-\tilde{y}^2\:.
\eeq
We again introduce the (continuous, but only piecewise smooth) function~$\alpha$ by
\beq \label{alpharel}
\alpha \;=\; {\mbox{Re}}\, \tilde{y}\:,
\eeq
define~$U$ by~\eqref{Udef} and (again for simplicity) choose~$W=U$.
We choose starting values for the functions~$\beta$ and~$R$ which are compatible with~\eqref{algebra}.
We can then apply Theorem~\ref{thm1} in various ways. If~$U<0$, i.e.\ $R+\beta$ and~$R-\beta$ have
opposite signs, depending on the sign of the determinator~$\mathfrak{D}$, we can apply
either part~(A) or part~(B). If however~$U \geq 0$, the functions $R+\beta$ and~$R-\beta$ are both positive,
and thus Theorem~\ref{thm1} applies only if~$\mathfrak{D}$ is positive, and in this case we
can apply both part~(A) and part~(B). If~$\beta$ becomes zero, it may be preferable to
switch to the simpler estimate of Theorem~\ref{thm2}, provided that the inequality~\eqref{invcond}
holds. In this way, we have different possibilities to obtain invariant disk estimates, and by
taking their intersection one gets even sharper estimates involving lens-shaped invariant regions.
Observe that on the boundaries between the WKB and Airy regions, the functions~$V$ and~$\alpha'$,
and consequently also~$U$, may have discontinuities. In order to satisfy~\eqref{algebra},
either~$R+\beta$ or~$R-\beta$ must ``jump'' discontinuously, in such a way that the new disk contains the old one.

The crucial point for making this procedure work is that one must be able to prescribe the sign
of~$\mathfrak{D}$ or to satisfy the inequality~\eqref{invcond}.
One method to achieve this is to modify the potentials~$V_{\mbox{\tiny{WKB}}}$ and~$V_{\mbox{\tiny{A}}}$ in~\eqref{VWKB} and~\eqref{VA}. The effect on~$\mathfrak{D}$ of modifying the potentials
can be seen most easily from Lemma~\ref{lemma34}. More specifically, in the WKB region,
the term~$2 \alpha\, {\mbox{Re}} (V-\tilde{V})$ in~\eqref{Dform}
can be suitably changed by modifying the real part
of~$V_{\mbox{\tiny{WKB}}}$, whereas the imaginary part of~$V_{\mbox{\tiny{WKB}}}$ affects
the term~$-\tilde{\beta} \,{\mbox{Im}}\, \tilde{V}$. Note that the derivative term in~\eqref{Dform}
as well as the derivative terms in~\eqref{tVdef} only give rise to small corrections.
In the Airy region, on the other hand, we can change the coefficients in the linear
function~$V_{\mbox{\tiny{A}}}(x)$ to modify the suitable terms in~\eqref{Dform}.
In order to satisfy the condition~\eqref{invcond}, it may be useful to apply the identity
\[ \eqref{invcond} \;=\; 2 \alpha\, {\mbox{\rm{Re}}} (V-\tilde{V}) \:+\:
\frac{1}{2}\: {\mbox{\rm{Re}}} (V-\tilde{V})' - \tilde{\beta}\: {\mbox{Im}}\, \tilde{V}
- |{\mbox{Im}} V| \sqrt{ {\mbox{Re}} (V-\tilde{V}) - \tilde{\beta}^2 } \]
(keeping in mind that in this case, $\beta \equiv 0$, and thus~$\tilde{\beta}$ should be small),
or else one can modify~$\alpha$, without respecting the relation~\eqref{alpharel}.

We now give two concrete examples to illustrate this method. In the first example, the
imaginary part of the potential is so small that the method works just as well for a real potential.
\begin{Example} \label{ex1} {\rm{
We consider the potential (inspired by the spheroidal
wave operator~\cite{FS})
\beq \label{Vdefex1}
V \;=\; 10000 \left( -\frac{1}{2} + (1 + 0.05 \,i) \sin^2 x \right) ,
\eeq
on the interval~$[0, \frac{\pi}{2}]$.
Due to the large prefactor, the WKB condition~\eqref{WKBcond}
is satisfied except in a small neighborhood of the point~$x=\frac{\pi}{4}$, where~$|V|$ is small.
Thus we divide our interval into three regions:
\begin{itemize}
\item[(a)] The ``classically allowed'' WKB region $[0, 0.715]$
\item[(b)] The Airy region $[0.715, 0.83]$ near the ``classical turning point''
\item[(c)] The  ``classically forbidden'' WKB region $[0.83, \frac{\pi}{2}]$
\end{itemize}
(The precise choice of the boundary points is arbitrary and has no major effect. Moreover, since the factor $0.05$
is so small, we simply adopted the terminology from quantum mechanics, disregarding the effect of the imaginary part of the potential.) In the Airy region we approximate the potential by its linear Taylor series around
the point~$x=\frac{\pi}{4}$. We glue together the corresponding WKB and Airy functions, starting at~$x=0$
with the WKB wave function~\eqref{VWKB} with~$V_{\mbox{\tiny{WKB}}}=V$ and choosing
the plus sign. In the region~(c) we also choose~$V_{\mbox{\tiny{WKB}}}=V$.
This gives the approximate wave function~$\tilde{\phi}$, and we define the corresponding
approximate Riccati solution by~$\tilde{y} = \tilde{\phi}'/\tilde{\phi}$. Following the general procedure
described earlier in this section, we choose~$\alpha$ according to~\eqref{alpharel} and set~$W=U$.
To better illustrate our estimates, we choose the radius~$R$ of the initial disk to be~$2.5$, although choosing
it equal to zero would give a better estimate. In Figure~\ref{fig4} (left), the imaginary part of
\begin{figure}%
\begin{center}%
 \includegraphics[width=7.2cm]{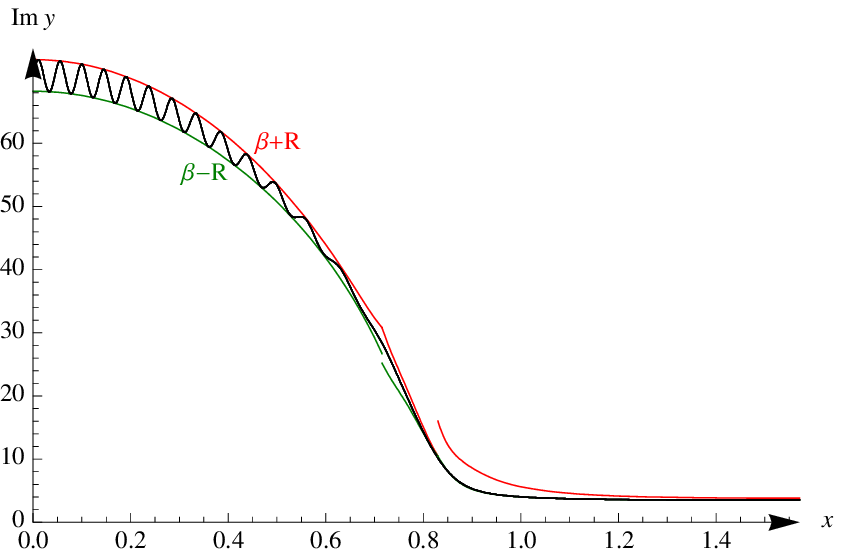}%
 \includegraphics[width=7.2cm]{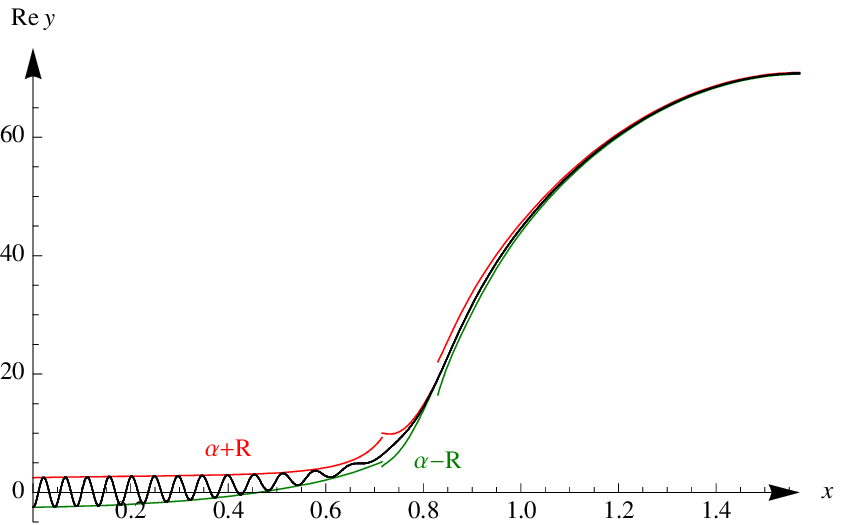}%
\caption{Upper and lower bounds for the imaginary (left) and real (right) parts of the Riccati solution
in Example~\ref{ex1}.}%
\label{fig4}%
\end{center}%
\end{figure}
an exact numerical solution~$y$ starting on the boundary of the initial circle is plotted together with the upper
bound~$\beta +R$ and the lower bound~$\beta-R$. Likewise, in Figure~\ref{fig4} (right), the real
part of the solution as well as the upper bound~$\alpha+R$ and the lower bound~$\alpha-R$ are given.
On sees that the invariant disks jump discontinuously at the glueing points, such that the new disk
contains the old disk. In Figure~\ref{fig5},

the invariant disks are plotted for discrete values of~$x$,
and the black dots denote the exact solution at these values of~$x$.
We point out that the invariant disk always stays in the upper half plane.
In region~(c), the determinator is negative, so that part (A) of Theorem~\ref{thm1}
applies. This means geometrically that the lower bound~$\beta-R$ approaches the exact solution
exponentially fast, whereas the upper bound~$\beta+R$ is not as good an approximation.

In order to get a better upper bound, we changed the WKB potential in region~(c) to
\beq \label{wkbnine}
V_{\mbox{\tiny{WKB}}} \;=\; 0.9\: V \:.
\eeq
This makes the determinator negative, so that part (B) of Theorem~\ref{thm1} applies.
Since~$V_{\mbox{\tiny{WKB}}}$ in~\eqref{wkbnine} deviates considerably from~$V$,
we cannot expect that our approximation will be good in region~(c). This is reflected
in Figure~\ref{fig6} by the fact that~$R$ becomes large if~$x$ approaches~$\frac{\pi}{2}$.
Nevertheless, this estimate is still useful because the upper bound~$\beta+R$ approaches the exact
solution exponentially fast.
Taking the intersection of the invariant disks with those from Figure~\ref{fig5}, one gets
lens-shaped invariant regions, thus estimating the exact solution up to exponentially decaying errors
(see Figure~\ref{fig7}). This exponential shrinkage of the invariant region can be understood by the
fact that the the potential is slowly varying, and that for a constant potential the solutions would tend exponentially fast to the stable fixed point $\sqrt{V}$ (see Section~\ref{sec2}).

Finally, in Figure~\ref{fig8} we show invariant disk estimates, where in region~(c) we applied
Theorem~\ref{thm2} and chose~$\beta \equiv 0$. This estimate is not as good as the previous estimate,
but on the other hand Theorem~\ref{thm2} is easier to apply, and the coarser estimate might be
sufficient for some applications. \QEDrem
}} \end{Example}

\begin{figure}%
\begin{center}%
\includegraphics[width=8cm]{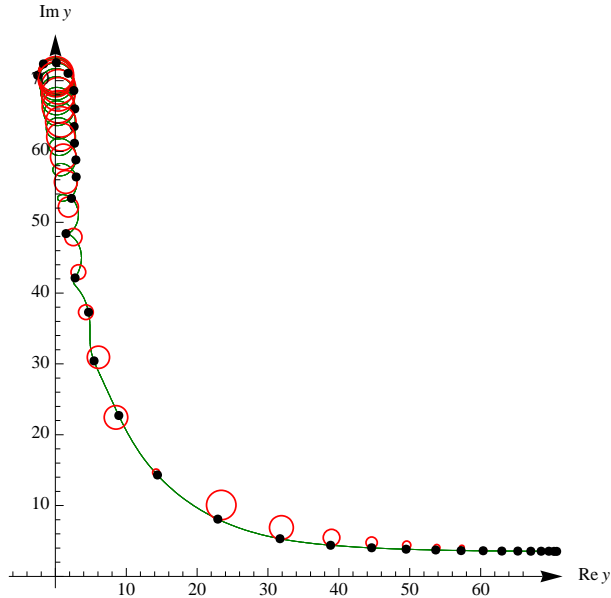}%
\caption{Invariant disks for the Riccati solution in Example~\ref{ex1}.}%
\label{fig5}%
\end{center}%
\end{figure}%

\begin{figure}%
\begin{center}%
\includegraphics[width=7.2cm]{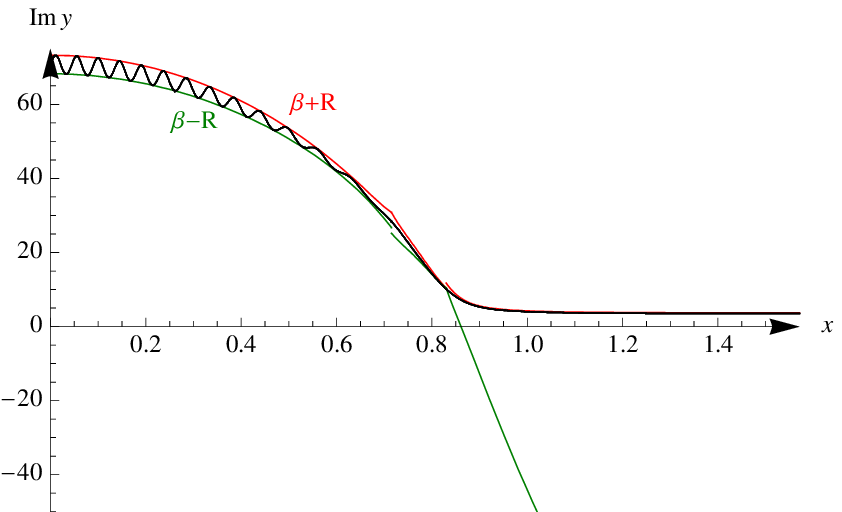}%
 \includegraphics[width=7.2cm]{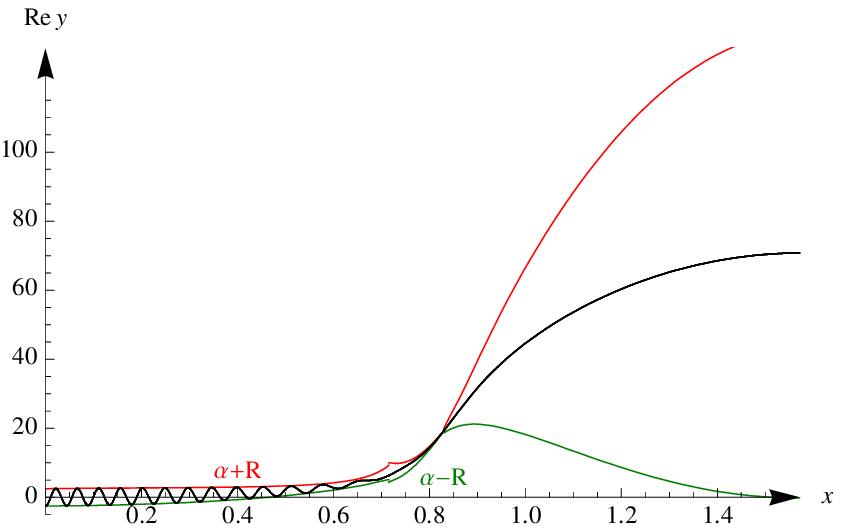}%
\caption{Upper and lower bounds after flipping the sign of~$\mathfrak{D}$ in Example~\ref{ex1}.}%
\label{fig6}%
\end{center}%
\end{figure}%

\begin{figure}%
\begin{center}%
\includegraphics[width=14cm]{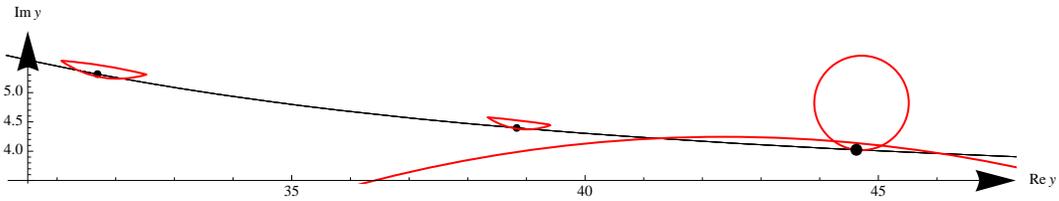}%
\caption{Lens-shaped region estimates for the Riccati equation in Example~\ref{ex1}.}%
\label{fig7}%
\end{center}%
\end{figure}%

\begin{figure}%
\begin{center}%
\includegraphics[width=7cm]{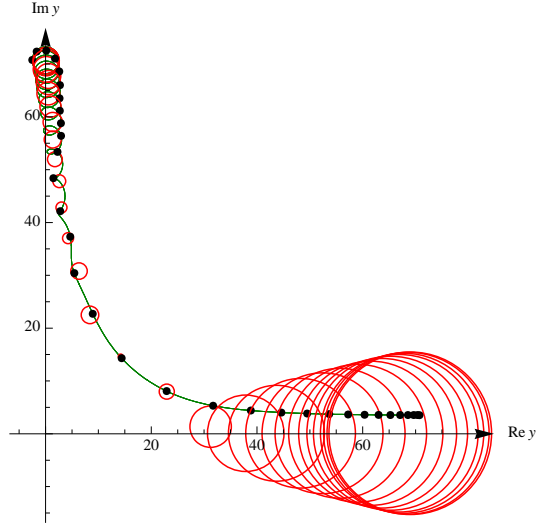}%
\caption{Invariant disk estimate applying Theorem~\ref{thm2} in Example~\ref{ex1}.}%
\label{fig8}%
\end{center}%
\end{figure}%

\begin{figure}[b]%
\begin{center}%
 \includegraphics[width=7.2cm]{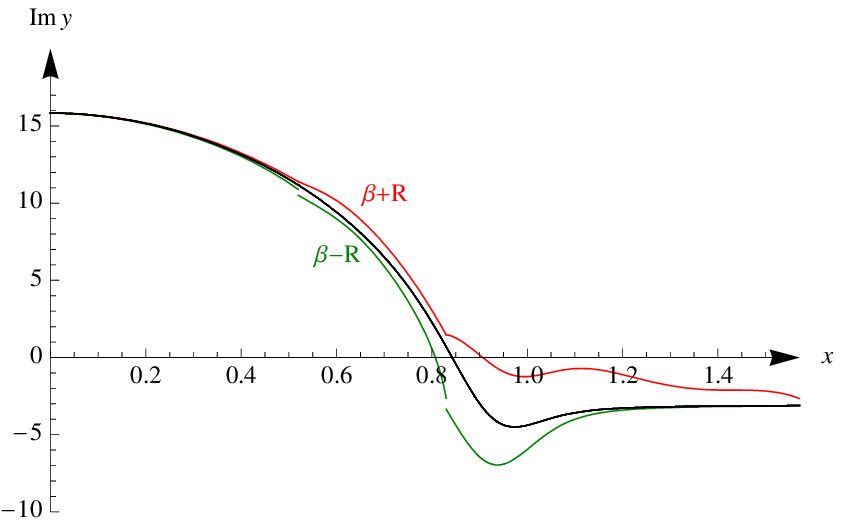}%
 \includegraphics[width=7.2cm]{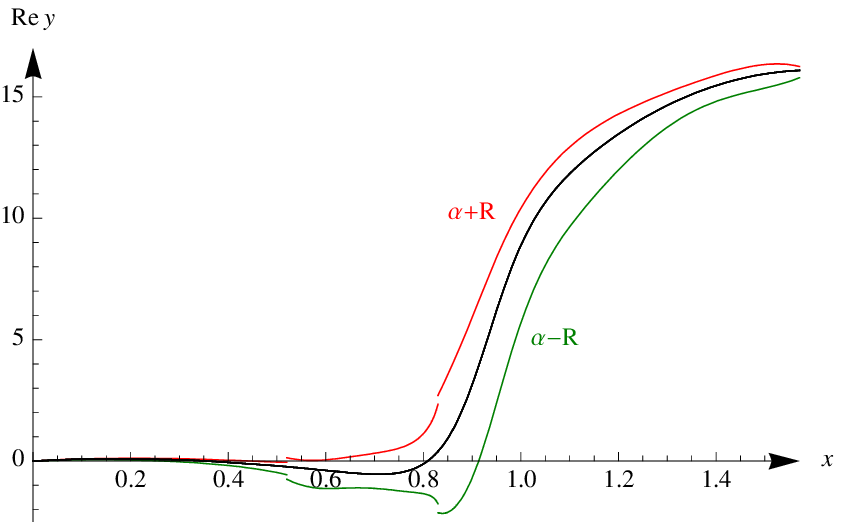}%
\caption{Upper and lower bounds for the imaginary (left) and real (right) parts of the Riccati solution
in Example~\ref{ex2}.}%
\label{fig9}%
\end{center}%
\end{figure}%

\begin{figure}[tbp]%
\begin{center}%
 \includegraphics[width=7.2cm]{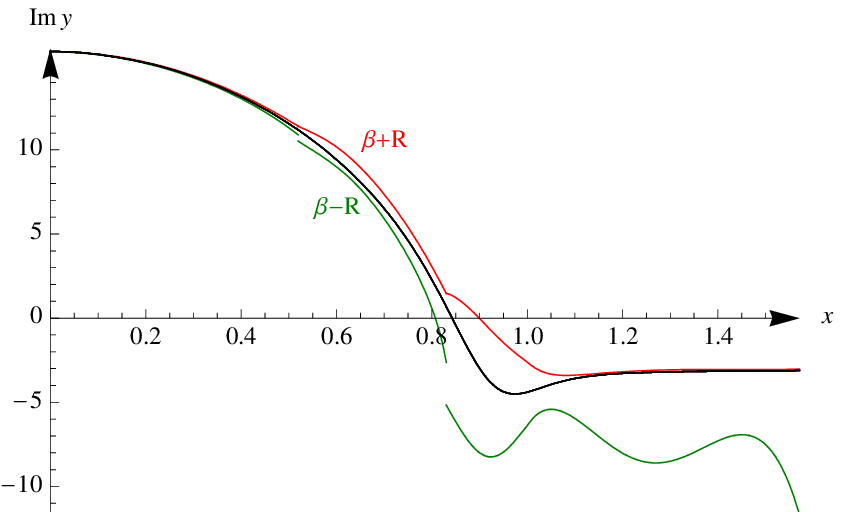}%
 \includegraphics[width=7.2cm]{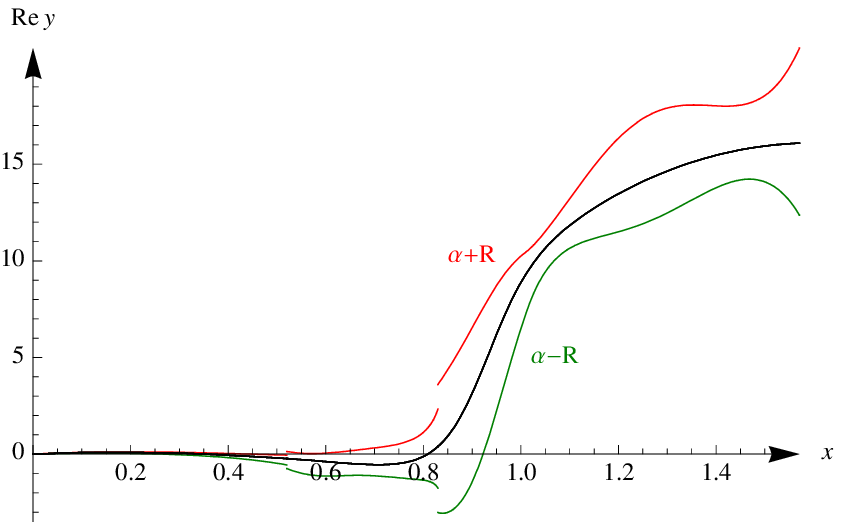}%
\caption{Upper and lower bounds after flipping the sign of~$\mathfrak{D}$ in Example~\ref{ex2}.}%
\label{fig10}%
\end{center}%
\end{figure}%

\begin{figure}[tbp]%
\begin{center}%
 \includegraphics[width=10cm]{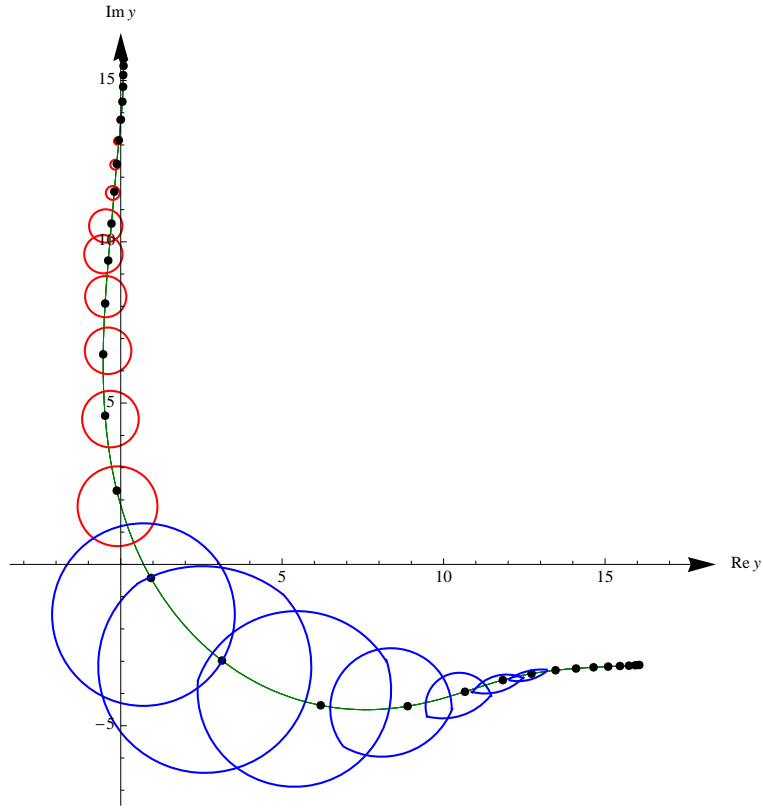}%
\caption{Invariant disk (red) and lens-shaped region estimates (blue) for the Riccati equation in Example~\ref{ex2}.}%
\label{fig11}%
\end{center}%
\end{figure}%

\begin{figure}[tbp]%
\begin{center}%
 \includegraphics[width=7.2cm]{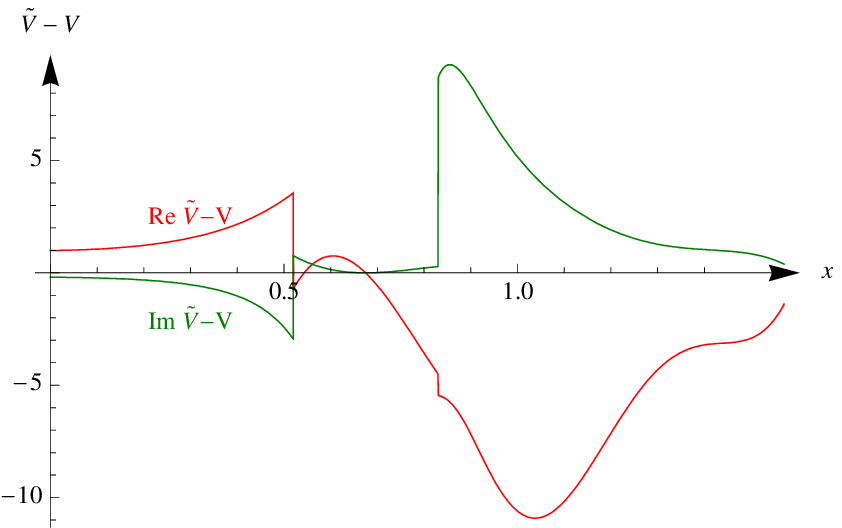}%
 \includegraphics[width=7.2cm]{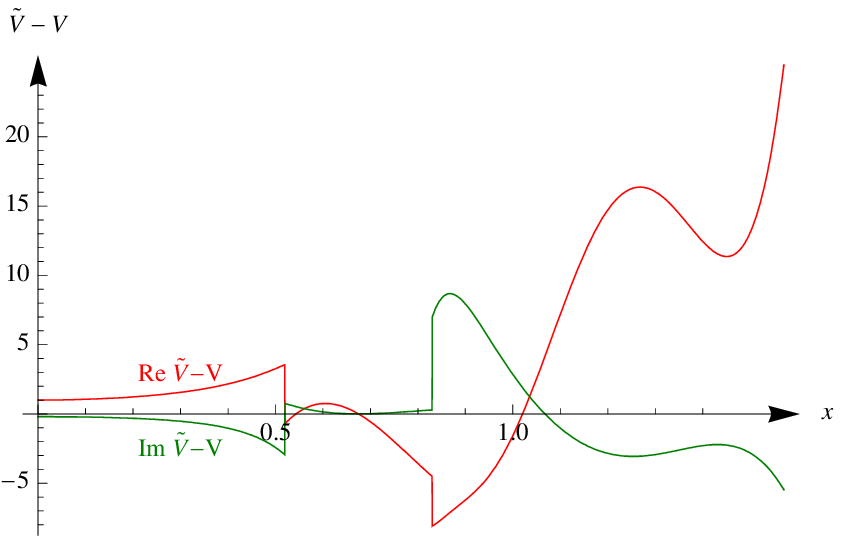}%
\caption{The potential~$\tilde{V}-V$ as used for the estimates in Figure~\ref{fig9} (left) and Figure~\ref{fig10} (right).}%
\label{fig12}%
\end{center}%
\end{figure}%

In the previous example, the semi-classical approximation was very good because of the  
large prefactor in~\eqref{Vdefex1}. In our next example, the WKB/Airy wave function is not such a good approximation, making the error estimates more subtle.
Moreover, the imaginary part of~$V$ has the opposite sign, thereby forcing the Riccati solution
to cross the real axis. By suitably adjusting the sign of the determinator, we will arrange that the
invariant disks also move from one half plane to the other.
\begin{Example} \label{ex2}
{\rm{ We consider on the interval~$[0, \frac{\pi}{2}]$ the potential
\[ V \;=\; 500 \left( -\frac{1}{2} + (1 - 0.2 \,i) \sin^2 x \right) . \]
We choose the WKB regions to be~$[0, 0.52]$ (region~(a)) and~$[0.83, \frac{\pi}{2}]$ (region~(c)),
and the intermediate region~$[0.52, 0.83]$ (region~(b)) is the Airy region.
In Figures~\ref{fig9} and~\ref{fig10} upper and lower bounds are given. In Figure~\ref{fig11} we plot
the corresponding invariant disk estimates as well as lens-shaped invariant regions obtained by
taking the intersection of the disks of Figures~\ref{fig9} and~\ref{fig10}.

Note that now the invariant disks cross the real axis. When this happens, one must ensure
when applying Theorem~\ref{thm1} that the denominator in~\eqref{Rpb} or~\eqref{Rpb2}
never becomes zero, because otherwise we would lose control of the estimates.
This means that if the function~$R-\beta$ vanishes, we must arrange the sign of the
determinator such that case~(A) applies, whereas~$R+\beta$ may vanish only when we are in case~(B).
To this end, the potentials~$V_{\mbox{\tiny{WKB}}}$ and~$V_{\mbox{\tiny{A}}}$ are chosen as follows:

In region~(a), we choose $V_{\mbox{\tiny{WKB}}}=V$. In region~(b), for~$V_{\mbox{\tiny{A}}}$
we took the linear Taylor polynomial and decreased the real part of the linear term so as to
make~$\mathfrak{D}$ positive. Then we are in case~(B) of Theorem~\ref{thm1}, in which the function~$\beta-R$ can cross
smoothly across the real line. Shortly after~$\beta-R$ has become negative, we switch to case~(A)
of Theorem~\ref{thm1}, so that~$\beta+R$ can smoothly flip sign. For the estimates of Figure~\ref{fig9}
we choose the potential $V_{\mbox{\tiny{WKB}}}=V$ in region~(c) such that~$\mathfrak{D}$
is always positive. Then~$\beta-R$ is a good approximation. For the estimates of Figure~\ref{fig10},
however, we modified~$V_{\mbox{\tiny{WKB}}}$ in region~(c) such as to arrange that~$\mathfrak{D}$ becomes negative, with the result that~$\beta+R$ is a good approximation.

Our choice of the approximate potentials is shown in Figure~\ref{fig12}, where we plot the
real and imaginary parts of~$\tilde{V}-V$, with~$\tilde{V}$ as in~\eqref{tVpot1} and~\eqref{tVpot2}.
We point out that the scale in these plots is one order of magnitude smaller than the scale of~$V \sim 250$.
This illustrates that slight modifications of the approximate potentials suffice to arrange the
appropriate sign of~$\mathfrak{D}$. The detailed form of the choice of the approximate potentials
affects the function~$\beta+R$ in Figure~\ref{fig9} (left) or the function~$\beta-R$ in
Figure~\ref{fig10} (left), as well as the functions~$\alpha \pm R$ on the right of these figures.
However, the corresponding lens-shaped invariant regions are insensitive
to the detailed choice of~$\tilde{V}$; see Figure~\ref{fig11}. \\ \hspace*{1cm} \QEDrem
}} \end{Example}

\noindent
{\em{Acknowledgments:}} 
We would like to thank the referees for valuable comments.
We are grateful to the Alexander-von-Humboldt Foundation as well as the
Vielberth Foundation, Regensburg, for their generous support.


\end{document}